# Smooth and non-smooth dependence of Lyapunov vectors upon the angle variable on a torus in the context of torus-doubling transitions in the quasiperiodically forced Hénon map

Alexey Yu. Jalnine[(a) *], Andrew H. Osbaldestin[(b)]

[(a)]*Institute of Radio-Engineering and Electronics of RAS, Saratov Division, Zelenaya 38, Saratov, Russia*
[(b)]*Department of Mathematics, University of Portsmouth, Portsmouth, PO1 3HE, UK*

**Abstract**

A transition from a smooth torus to a chaotic attractor in quasiperiodically forced dissipative systems may occur after a finite number of torus-doubling bifurcations. In this paper we investigate the underlying bifurcational mechanism which seems to be responsible for the termination of the torus-doubling cascades on the routes to chaos in invertible maps under external quasiperiodic forcing. We consider the structure of a vicinity of a smooth attracting invariant curve (torus) in the quasiperiodically forced Hénon map and characterize it in terms of Lyapunov vectors, which determine directions of contraction for an element of phase space in a vicinity of the torus. When the dependence of the Lyapunov vectors upon the angle variable on the torus is smooth, regular torus-doubling bifurcation takes place. On the other hand, the onset of non-smooth dependence leads to a new phenomenon terminating the torus-doubling bifurcation line in the parameter space with the torus transforming directly into a strange nonchaotic attractor. We argue that the new phenomenon plays a key role in mechanisms of transition to chaos in quasiperiodically forced invertible dynamical systems.

## 1. Introduction

The investigation of transition mechanisms from quasiperiodic dynamics to chaos is one of the central topics in contemporary nonlinear science. Starting with the classic works of Landau [1] and Ruelle and Takens [2], many researchers have undertaken theoretical [3-9] and experimental [10-13] studies of this problem. As is well-known, an image of a regular quasiperiodic motion in the phase space of a dissipative dynamical system is a smooth attracting *ergodic torus*. One convenient way to investigate mechanisms for the destruction of an ergodic torus is to consider quasiperiodically forced systems: in such systems the frequency ratios appear as independent parameters, and can be effectively controlled both in numerics and in experiments. Quasiperiodically forced systems have become popular models for studies of the transition from quasiperiodicity to chaos after the discovery of a *strange nonchaotic attractor* (SNA) by Grebogi, Pelican, Ott and Yorke in 1984 [14]. An SNA typically appears in the intermediate region between order and chaos and possesses a mixture of features of regular and chaotic attractors. Attractors of this type are nonchaotic in the sense that only nonpositive Lyapunov exponents occur, but they possess a fractal-like geometrical structure, which justifies the term "strange". (For more details on structure and properties of SNA, see Refs. [15-22].)

---
[*] Corresponding author. E-mail address: chaos777@rol.ru

One of the important observations, made in the 1980s by Anishchenko [8] and Kaneko [9], is that the destruction of a smooth torus and the appearance of chaos may be preceded by a finite number of torus-doubling bifurcations. Therefore, much attention is focused on numerical [23-31] and experimental [32,33] studies of dynamical transitions in period-doubling systems under the effect of an external quasiperiodic force. When the amplitude of the external quasiperiodic force is fixed and the nonlinearity parameters of the model system are varied, a sequence of torus-doubling bifurcations can occur. Such a sequence is typically terminated by the onset of an SNA, followed by a further transition to chaos. The number of torus-doubling bifurcations in the sequence depends upon the amplitude of the external quasiperiodic force. For the case of sufficiently large amplitudes, a simple smooth torus may transform into an SNA. For small amplitude values, several torus-doubling bifurcations may occur before the SNA arises. The number of torus-doubling bifurcations grows as the amplitude of the quasiperiodic force is decreased. However, this number appears to be finite for any fixed nonzero amplitude. (See numerical results presented in Ref. [24].) An infinite bifurcation sequence can occur only for the case of the driving force amplitude equal to zero, as follows from the analysis developed in Ref. [23]. Thus an important issue is to understand the reason for the termination the torus-doubling cascades in the quasiperiodically forced systems.

For non-invertible unimodal maps the mechanism of termination of torus-doubling cascades appears to be closely associated with the critical behavior studied by Kuznetsov et al. [26]. The line of torus-doubling bifurcation in the parameter space of the quasiperiodically forced logistic map terminates at a special critical point, called the Torus Doubling Terminal (TDT). (The corresponding values of the quasiperiodic force amplitude and the nonlinearity parameter will hereafter be referred to as the critical parameter values.) The termination of the bifurcation line is associated with the tangency of the attractor with the line of zero derivative of the map. This event changes the character of the bifurcation, which becomes phase-dependent, and the attractor of the system becomes non-smooth. For amplitudes of the quasiperiodic force above the critical value the sign of the derivative depends upon the angle variable on the torus, therefore, regular torus-doubling bifurcation becomes impossible. Numerical analysis shows that for small amplitudes of the quasiperiodic force a similar mechanism terminates the lines of doubling bifurcations for doubled, quadrupled, and other tori of this system [34]. Thus we can conclude that non-invertibility plays the role of a "terminator" for the torus-doubling cascades on the route to chaos in the quasiperiodically forced logistic map as well as for other noninvertible 1D maps of the same universality class.

It appears that the structure of the parameter space described above occurs in different period-doubling systems under external quasiperiodic forcing. For example, analogous transitions were observed in numerical experiments on a nonlinear dissipative oscillator under external two-frequency driving with irrational frequency ratio [35]. The Poincaré map in the phase space of such an oscillator is a smooth invertible 3D map with one quasiperiodic variable. The most-widely known example of such kind is a quasiperiodically forced Hénon map [30,31]. A smooth closed invariant curve (torus) in the phase space of this map corresponds to the Poincaré section of the torus in the phase space of biharmonically forced oscillator. Note that a reduction of the invertible 2D Hénon map in the limit of strong dissipation produces a noninvertible 1D logistic map. On the other hand, for dynamical systems determined by differential equations or for invertible maps, the mechanism of termination of the torus-doubling cascades obviously must be different from the above mentioned loss of invertibility, which works only for non-invertible forced 1D maps.

In order to understand the underlying mechanism of termination of the torus-doubling cascades in invertible systems, we consider in this paper the Hénon map driven by an external quasiperiodic force with an irrational frequency parameter, chosen to be the inverse golden mean. Since the torus-doubling bifurcation is local, we focus attention on a study of the vicinity of a smooth attracting invariant curve (torus) in this system. Such a vicinity can be characterized in terms of Lyapunov vectors, which determine directions of contraction for an element of phase volume around the attracting torus. The values of the Lyapunov vectors depend upon the angle variable on torus. If the dependence of the Lyapunov vectors upon the angle variable is smooth, torus-doubling bifurcation is possible. Alternatively, we observe a new transition, associated with the onset of non-smooth dependence of Lyapunov vectors upon the angle variable on the torus. We show that the latter transition prevents a torus-doubling bifurcation and terminates the line of this bifurcation in the phase space. Therefore, further evolution of the torus under variation of the parameters of the system is associated with the appearance of an SNA or of a chaotic transient. We argue that an analogous mechanism may be responsible for the prevention of doubling bifurcations for doubled, quadrupled, and other tori of the model system.

The paper is organized as follows. In *Section 2* we define Lyapunov vectors for quasi-periodic trajectories on a torus, and use them for a description of the mechanism of torus-doubling bifurcation. In *Section 3* we present numerical data and discuss smooth and non-smooth dependences of Lyapunov vectors upon the angle variable for different parameter values of the model system. In *Section 4* we explain the mechanism which prevents the torus-doubling bifurcation from the viewpoint of the method of rational approximation [15]. In the *Conclusion* we discuss the role of the new phenomenon associated with the appearance of non-smooth dependences of the Lyapunov vectors upon the angle variable on the torus in a general picture of transitions from quasiperiodicity to chaos, which involve different bifurcations of tori.

## 2. Characterization of the torus vicinity: Lyapunov vectors and invariant 2D manifolds

Let us start with an autonomous Hénon map:

$$x_{n+1} = a - x_n^2 + y_n$$
$$y_{n+1} = bx_n$$
(1)

where $0<b<1$. Let $(x_0,y_0)$ be a fixed point of this map. The multipliers of the fixed point are defined as $\mu_{1,2} = (S \pm \sqrt{S^2 - 4J})/2$, where $J = -b$ is a determinant of the Jacobi matrix of the map (1) and $S = -2x_0$ is a trace of this matrix at the fixed point $(x_0,y_0)$. Due to our choice of $b$, the condition holds $S^2-4J>0$. The last condition implies that the fixed point possesses two different real multipliers $\mu_{1,2}$ ($\mu_1\mu_2 = -b$), and, hence, the point is either a saddle or a stable node. For definiteness, let us suppose that $|\mu_1| > |\mu_2|$.

In the case of the saddle point ($|\mu_1| > 1$, $|\mu_2| < 1$), there are two invariant 1D manifolds (stable and unstable ones), which are represented by smooth invariant curves in the phase plane (see fig.1a). The two eigenvectors $\mathbf{k}^{1,2}$ of the Jacobian matrix (Lyapunov vectors) give the directions tangent to the invariant manifolds at the fixed point.

When $|\mu_{1,2}| < 1$, the fixed point is a stable node. In this case also we can define two Lyapunov vectors, which determine the directions of contraction for an element of phase space in a vicinity of the nodal fixed point. The *leading* vector $\mathbf{k}^1$, associated with the multiplier of largest modulus is tangent to the set of stable invariant manifolds, as shown in fig.1b. (See also Ref. [36].) The vector $\mathbf{k}^2$, referred to as the *non-leading* eigenvector, is tangent to the single non-leading stable invariant manifold.

Now we modify the map (1) by adding an external quasiperiodic force, and consider the model map in $\mathbf{R}^2 \times \mathbf{T}^1$:

$$x_{n+1} = a - x_n^2 + y_n + \varepsilon \cos 2\pi \theta_n,$$
$$y_{n+1} = bx_n, \qquad (2)$$
$$\theta_{n+1} = \theta_n + \omega, \bmod 1,$$

where $\omega$ is an irrational number, which we set equal to the inverse golden mean: $\omega = (\sqrt{5}-1)/2$. For $\varepsilon = 0$ map (2) has a trivial invariant curve (torus)

$$T_0: \{(x, y, \theta) \in \mathbf{R}^2 \times \mathbf{T}^1 \mid x = x_0, y = y_0, \theta \in [0,1)\}.$$

Obviously, in this case a structure of a vicinity of the torus $T_0$ will be determined by multipliers of the fixed point $(x_0, y_0)$.

If $|\mu_1| > 1$ and $|\mu_2| < 1$, the torus $T_0$ is of a saddle type, and there are two invariant manifolds, unstable and stable, which we denote as $W^u$ and $W^s$, respectively. The manifolds are represented by smooth 2D surfaces in the 3D phase space, as shown in fig.1c. At any point of the saddle torus one can define two directions, which are tangent to the invariant manifolds and orthogonal to the axis of the angle variable $\theta$. For $\varepsilon = 0$ these directions are given simply by the Lyapunov vectors $\mathbf{k}^{1,2}$ of the fixed point $(x_0, y_0)$ of the map (1).

Likewise, if $|\mu_{1,2}| < 1$, the torus $T_0$ is of a stable nodal type, and, again, at any point of a stable nodal torus one can define two Lyapunov vectors, which determine two directions of contraction for an element of phase space in vicinity of the torus. The rate of contraction in each direction is characterized by the respective Lyapunov exponent ($\sigma_{1,2} = \log|\mu_{1,2}|$). If we introduce 2D stable invariant manifolds associated with the nodal torus (as extension of 1D invariant manifolds of nodal fixed point of the map (1)), then two Lyapunov vectors $\mathbf{k}^{1,2}$ will define two directions tangent to the manifolds and orthogonal to the axis of angle variable $\theta$ (see fig.1d). The leading vector $\mathbf{k}^1$ is tangent to a continuum of stable 2D manifolds (we arbitrarily choose one of them and refer it to as $W^1$), while the vector $\mathbf{k}^2$ is tangent to one special non-leading stable manifold $W^2$. The remainder of this article is concerned with the stable nodal torus and its vicinity[1].

Now let $\varepsilon \neq 0$. For typical values of $a$ and $b$ apart from the bifurcational points of the map (2), a small quasiperiodic perturbation will not destroy the torus and the smooth 2D manifolds. Thus, for small nonzero $\varepsilon$ the map (2) possesses a nontrivial torus:

$$\mathrm{T}: \{(x, y, \theta) \in \mathbf{R}^2 \times \mathbf{T}^1 \mid x = x(\theta), y = y(\theta), \theta \in [0,1]\}; \qquad (3)$$

the stable 2D manifolds $W^{1,2}$ in a vicinity of the torus T become distorted, but remain smooth 2D surfaces. The Lyapunov vectors, which are tangent to the manifolds and orthogonal to the $\theta$-axis, now depend on the angle variable $\theta$: $\mathbf{k}^{1,2} = \mathbf{k}^{1,2}(\theta)$. While the manifolds are smooth, the vector-functions $\mathbf{k}^{1,2}(\theta) = (k_x^{1,2}(\theta), k_y^{1,2}(\theta), 0)$ remain differentiable. As the parameter $\varepsilon$ increases (other parameters of the map (2) we suppose to be fixed), the plots of the functions $k_{x,y}^{1,2}(\theta)$ may become more and more distorted, until these functions lose differentiability at some critical value of $\varepsilon$. Appearance of non-smooth dependences of Lyapunov vectors $\mathbf{k}^{1,2}$ upon the angle variable $\theta$ apparently provides an evidence of the destruction of the smooth 2D manifolds in a vicinity of the torus T.

Let us discuss a role of Lyapunov vectors and 2D invariant manifolds in the mechanism of the torus-doubling bifurcation in the map (2). On the threshold of bifurcation, the map possesses a nodal torus T, shown in fig.1d. As a control parameter of the system passes through the bifurcational value, the nodal torus T looses stability and becomes of a saddle type. The loss of stability of the torus T occurs along the less stable leading direction $\mathbf{k}^1(\theta)$, as the corresponding Lyapunov exponent $\sigma_1$ passes through zero. A pair of smooth curves 2T ("double torus") appears in a vicinity of T; a trajectory on the double torus visits two curves alternately. The "leading" manifold $W^1$ of the parent nodal torus T transforms after bifurcation into the unstable manifold $W^u$ of the saddle torus T. The newly-born double torus 2T belongs to the smooth manifold $W^u$, as shown in fig.1e. Note that the vector-function $\mathbf{k}^1(\theta)$ determines the direction tangent to $W^u$. Hence, immediately after the bifurcation the vector-function $\mathbf{k}^1(\theta)$ determines in linear approximation the direction from the saddle torus T to the newly-born double torus 2T. Since all the tori (T and 2T) are smooth, and they belong to the smooth manifold $W^u$, the dependence $\mathbf{k}^1(\theta)$ will be also smooth. On the other hand, the non-smooth dependence of $\mathbf{k}^1(\theta)$ upon $\theta$ would imply that a newly-born object (born instead of 2T) must also be non-smooth as it belongs to a non-smooth manifold $W^u$. Thus, existence a smooth vector-function $\mathbf{k}^1(\theta) = (k_x^1(\theta), k_y^1(\theta), 0)$ appears to be a necessary condition for a possibility of the regular torus-doubling bifurcation. The loss of smoothness of the dependence $\mathbf{k}^1(\theta)$ provides us with evidence that torus-doubling bifurcation becomes impossible. Let us consider now the methods for numerical computation of the dependences $\mathbf{k}^{1,2}(\theta)$ and for the analysis of their smoothness.

---

[1] Note that, besides stable nodes and saddles, a dissipative map may possess a fixed point of focal type, which is characterized by complex conjugate multipliers ($\mu_1 = \mu_2^*$). In such case addition of the quasiperiodic variable $\theta$ gives a smooth torus that has a vicinity of focal type. The Lyapunov vectors are not defined in the focus. Therefore, the 2D invariant manifolds turn around the stable torus of focal type. In fact, the one time iterated Hénon map (1) does not possess focal fixed points at $b > 0$. However, it has stable periodic orbits of periods $2^n$, $n \geq 2$, which are characterized by complex values of $\mu_{1,2}$. Further we will observe some quasiperiodic regimes arising from focal periodic orbits, although they do not play a significant role in the present work.

First, let us turn to a case when the functions $\mathbf{k}^{1,2}(\theta)$ are smooth. Let there be a point $(x_0, y_0, \theta_0)$, which belongs to the torus (3). In order to define the Lyapunov vectors $\mathbf{k}^{1,2}(\theta_0)$ at this point we iterate map (2) starting from $(x_0, y_0, \theta_0)$ and obtain an orbit: $(x_0, y_0, \theta_0)$, $(x_1, y_1, \theta_1)$, ..., $(x_n, y_n, \theta_n)$. Let a vector $\mathbf{k}_0$ be collinear to the vector $\mathbf{k}^1(\theta_0)$ (or $\mathbf{k}^2(\theta_0)$) at the initial point. After one iteration of map (2), this vector will be mapped into the vector $\mathbf{k}_1$, which is collinear to the vector $\mathbf{k}^1(\theta_1)$ (or $\mathbf{k}^2(\theta_1)$) at the point $(x_1, y_1, \theta_1)$. The evolution of $\mathbf{k}_0$ is described by the Jacobi matrix $\hat{J}(x_0, y_0, \theta_0)$ of the map (2):

$$\mathbf{k}_1 = \hat{J}(x_0, y_0, \theta_0)\mathbf{k}_0. \tag{4}$$

After $n$ iterations the operator $\hat{J}^{(n)}$ of evolution of the vector is

$$\hat{J}^{(n)} = \hat{J}(x_{n-1}, y_{n-1}, \theta_{n-1})\hat{J}(x_{n-2}, y_{n-2}, \theta_{n-2})\ldots\hat{J}(x_0, y_0, \theta_0).$$

Thus, we obtain a sequence of vectors $\mathbf{k}_1$, $\mathbf{k}_2$, ..., $\mathbf{k}_n$, with $\mathbf{k}_n = \hat{J}^{(n)}\mathbf{k}_0$, which are collinear to the Lyapunov vectors at the respective points of the orbit. Now, in order to define the initial vector $\mathbf{k}_0$, we consider a subsequence of the trajectory points which converges to the initial point $(x_0, y_0, \theta_0)$. Since we have chosen $\omega$ equal to the inverse golden mean we take the subsequence $(x_{F_0}, y_{F_0}, \theta_{F_0}), \ldots, (x_{F_k}, y_{F_k}, \theta_{F_k})$, where $F_k = 1, 2, 3, 5, 8, 13\ldots$ are the Fibonacci numbers. Under the assumption of smoothness of $\mathbf{k}^{1,2}(\theta)$, the sequence of vectors $\mathbf{k}_{F_0}, \ldots, \mathbf{k}_{F_k}$ also converges to the vector $\mathbf{k}_0$ at the initial point. Hence, we come to a conclusion, that

$$\mathbf{k}_{F_k} = \hat{J}^{(F_k)}\mathbf{k}_0 \to \mu_{F_k}\mathbf{k}_0 \quad \text{as } k \to \infty, \tag{5}$$

where $\mu_{F_k}$ is a some coefficient. Thus, we obtain an eigenvalue problem for the matrix

$$\hat{J}^{(F_k)} = \begin{bmatrix} J_{11} & J_{12} & J_{12} \\ J_{21} & J_{22} & J_{23} \\ 0 & 0 & 1 \end{bmatrix}.$$

One of the eigenvectors of the matrix $\hat{J}^{(F_k)}$ corresponds to a trivial unit eigenvalue associated with the angle variable. The other two eigenvectors have the form $\mathbf{m}_{F_k}^{1,2} = (m_x^{1,2}, m_y^{1,2}, 0)$, orthogonal to the axis of the angle variable. Hence, at the point $(x_0, y_0, \theta_0)$ one can define two Lyapunov vectors $\mathbf{k}^{1,2}(\theta_0)$ as the limits for eigenvectors $\mathbf{m}_{F_k}^{1,2}$ at $k \to \infty$. Analogous arguments can be developed for any point $(x, y, \theta)$ of the torus (3). Note that the relation (5) makes it possible to determine two nontrivial Lyapunov exponents for a quasiperiodic trajectory on the torus as

$$\sigma_{1,2} = \lim_{k \to \infty} (1/F_k) \ln | \mu_{F_k}^{1,2} |.$$

In the limit $k \to \infty$ the values of $\sigma_{1,2}$ do not depend on the initial phase $\theta_0$ and characterize the entire torus, since the quasiperiodic trajectory fills the torus densely due to ergodicity of the quasiperiodic motion.

In practice, the method of definition of the Lyapunov vectors described above is inconvenient for numerical computations. Moreover, the method was based on an assumption of differentiability of $\mathbf{k}^{1,2}(\theta)$. On the other hand, we should take into account that such dependences can be either differentiable or non-differentiable. Nevertheless, due to a possibility of definition of $\mathbf{k}^{1,2}$ as the eigenvectors of an operator (see (5)), we can suggest another simple way for their determination.

Let us suppose that the vector functions $\mathbf{k}^{1,2}(\theta)$ corresponding to the leading and non-leading Lyapunov vectors are normalized to unity at any point of the torus (3). We can consider the evolution of an *arbitrarily* chosen vector $\mathbf{k}_0 = (k_{x,0}, k_{y,0}, 0)$ along the trajectory $(x_0,y_0,\theta_0)$, $(x_1,y_1,\theta_1)$, …, $(x_n,y_n,\theta_n)$ under iterations of the linearized map (4). Multiplying by the Jacobian matrix at each point of the trajectory, and than normalizing, we obtain the following map:

$$\mathbf{k}'_{n+1} = \hat{J}(x_n, y_n, \theta_n)\mathbf{k}_n,$$
$$\mathbf{k}_{n+1} = |\mathbf{k}'_{n+1}|^{-1}\mathbf{k}'_{n+1}, \qquad (6)$$
$$\theta_{n+1} = \theta_n + \omega, \mod 1.$$

As we know, in a typical case after a sufficiently large number of iterations, an arbitrarily assigned vector tends to the direction corresponding to the largest Lyapunov exponent (e.g. Ref. [37]). Since we have chosen $\mathbf{k}_0$ initially orthogonal to the phase axis, this direction will be given by the leading Lyapunov vector $\mathbf{k}^1(\theta)$. Thus, $\mathbf{k}_n$ tends to $\pm\mathbf{k}^1(\theta_n)$ as $n \to \infty$. A plot of the function $\pm\mathbf{k}^1(\theta)$ may be interpreted as an image of the attractor of the map (6). Note that for any quasiperiodic trajectory on the torus (3) the values $x_n$ and $y_n$ are functions of the angle variable $\theta_n$: $x_n = x(\theta_n)$, $y_n = y(\theta_n)$. This fact makes it possible to consider the map (6) as a usual quasiperiodically forced map and allows us to use standard methods for the analysis of its dynamical regimes. For instance, to obtain the "leading" Lyapunov vector $\mathbf{k}^1(\theta_0)$ at the point $(x_0, y_0, \theta_0)$ on the torus, we should start iterating (6) from the initial angle $\theta_{-n}$ [$= \theta_0 - n\omega$, mod 1], where $n$ is sufficiently large, with an arbitrarily chosen initial condition $\mathbf{k}_{-n}$.

Now let us consider possible types of attractors of the map (6). In the context of further numerical analysis, the following three cases appear to be essential:

(C1) the map has two attractors represented by smooth invariant curves $\pm\mathbf{k}^1(\theta)$, which are symmetric with respect to the axis of angle variable $\theta$;

(C2) the map has one attractor, which consists of two smooth curves $\pm\mathbf{k}^1(\theta)$, visited alternately at iterations of the map;

(C3) the attractor of the map is *strange nonchaotic,* represented by the non-smooth[2] and double-valued function $\pm \mathbf{k}^1(\theta)$.

In the cases (C1) and (C2) the function $\mathbf{k}^1(\theta)$ is differentiable. It implies a smooth character of the dependence of leading Lyapunov vector upon the angle variable on the torus (3). The appearance of a strange nonchaotic attractor in the map (6) (case (C3)) provides evidence of a loss of differentiability of the vector-function $\mathbf{k}^1(\theta)$. Hence, in the last case the dependence of the leading Lyapunov vector upon the angle variable is non-smooth[3].

In the same way, we can determine the non-leading Lyapunov vector $\mathbf{k}^2(\theta)$, which corresponds to the second nontrivial Lyapunov exponent. For this, we invert the map (2) and consider an evolution of some arbitrary chosen vector $\mathbf{k}_0$ under iterations of the inverse map along the quasiperiodic trajectory on the torus (3), i.e. backward in time. Taking into account a normalization of the vector, we represent the evolution map as:

$$\begin{aligned} \mathbf{k}'_{n+1} &= \hat{\boldsymbol{J}}^{-1}(x_n, y_n, \theta_n)\mathbf{k}_n, \\ \mathbf{k}_{n+1} &= \left|\mathbf{k}'_{n+1}\right|^{-1}\mathbf{k}'_{n+1}, \\ \theta_{n+1} &= \theta_n - \omega, \bmod 1. \end{aligned} \qquad (7)$$

Here $\hat{\boldsymbol{J}}^{-1}(x,y,\theta)$ is the Jacobian matrix of the map inverse to the quasiperiodically forced Hénon map (2). Since the maps (6) and (7) are inverse with respect to each other, they possess identical invariant sets. Note, that the attracting invariant set of (6) (defined as $\pm \mathbf{k}^1(\theta)$) is a repellor for the map (7), while the attractor of the map (7) (given by $\pm \mathbf{k}^2(\theta)$) appears to be the repelling invariant set of the map (6). Hence, under iterations of (7) the vector $\mathbf{k}_n$ will tend to $\pm \mathbf{k}^2(\theta_n)$ as $n \to \infty$.

Thus, the problem of analysis of the dependences of "leading" and "non-leading" Lyapunov vectors $\mathbf{k}^{1,2}$ upon the angle variable $\theta$ is reduced to analysis of the attractors of the maps (6) and (7) in the space of Lyapunov vectors. Smoothness of the attractors represented by vector functions $\pm \mathbf{k}^{1,2}(\theta)$ implies smoothness of the dependences of Lyapunov vectors on the torus (3) upon the angle variable. The onset of strange nonchaotic attractors in the maps (6) and (7) indicates the loss of smoothness of the dependences of Lyapunov vectors $\mathbf{k}^{1,2}$ upon the angle variable $\theta$ and, as a consequence, the destruction of smooth 2D invariant manifolds in a vicinity of the nodal torus (3).

### 3. The loss of differentiability of the dependence of Lyapunov vectors upon the angle variable

Let us fix $b=0.5$ and $\varepsilon=0.6$, and consider the evolution of the attractor of the map (2) and the attractors of (6) and (7) in the Lyapunov space under variation of the parameter *a*.

---

[2] According to the results of Stark (see Ref. [19]), a SNA can not be the graph of a continuous function. Strictly speaking, the function $\pm \mathbf{k}^1(\theta)$ must be non-smooth and upper/lower semi-continuous.

[3] In a general case, we should consider one more possibility: (C4) the attractor of the map (6) represents a 3-frequency torus. In this case the vector-function $\mathbf{k}^1(\theta)$ cannot be defined. This situation takes place when the quasiperiodic forcing is added to a system with a focal fixed point. As we has already mentioned, the Hénon map does not possess such points for *b*>0. However, the 3-frequency quasiperiodic regime may be observed in the system (6) when we investigate the structure of vicinity of a double torus 2T or quadruple torus 4T of the map (2).

At *a*=0.55 the attractor of the map (2) is a smooth torus (fig.2a). The attractor of the map (6) is a double torus, i.e. it is represented by a pair of smooth curves $\pm \mathbf{k}^1(\theta)$, which image into each other at iterations of the map. The resulting plot of the function $k_x^1(\theta)$ is presented in fig.2b. The map (7) possesses two attracting invariant tori ($\pm \mathbf{k}^2(\theta)$), which are symmetric with respect to the axis of angle variable $\theta$. The plot of the function $k_x^2(\theta)$ is shown in fig.2c. One can see that both functions $k_x^{1,2}(\theta)$ are smooth: the Lyapunov vectors depend smoothly upon the angle variable $\theta$. As *a* is increased, the smooth vector functions $\pm \mathbf{k}^{1,2}(\theta)$ corresponding to attractors of the maps (6) and (7) become more and more distorted at small scales, until strange nonchaotic attractors arise simultaneously in (6) and (7) at the critical value $a_c \simeq 0.559$. The plots of the functions $k_x^{1,2}(\theta)$ at *a*=0.559 are presented in figs.2d,e. Thus, the dependences of the Lyapunov vectors upon the angle variable become non-differentiable. Note that the attractor of the map (2) still remains a smooth torus, as shown in fig.2f. The transition to SNA in this map (2) occurs only at $a_f \simeq 0.656$.

A smooth torus characterized by non-smooth dependences $\mathbf{k}^{1,2}(\theta)$ can be observed for all values of the parameter *a* within the interval $a \in [a_c, a_f)$. Numerical analysis shows that besides this interval there are other intervals of *a* with non-smooth dependences of the Lyapunov vectors upon the angle coordinate: $a \in [0.266, 0.416]$ and $a \in [0.484, 0.521]$. However, we emphasize that this is the first interval $[a_c, a_f)$, which is important for the explanation of the direct transition from smooth torus to SNA without torus-doubling bifurcation in the system (2). Non-smooth dependence of the Lyapunov vectors upon the angle variable for $a \in [a_c, a_f)$ makes the torus-doubling bifurcation impossible, and in this case the smooth torus directly transforms into the SNA via a gradual fractalization (as described in Ref. [25]).

Now let us consider the process of destruction of smooth dependences $\mathbf{k}^{1,2}(\theta)$ in some details. For this purpose we need to calculate the angle $\varphi(\theta)$ between "leading" and "non-leading" Lyapunov vectors $\mathbf{k}^{1,2}$ on the torus as a function of $\theta$. Since we have chosen $|\mathbf{k}^{1,2}(\theta)|=1$, we immediately get

$$\varphi(\theta) = \arccos(\mathbf{k}^1(\theta) \mathbf{k}^2(\theta)).$$

Then we take the least of the two angles: $\varphi$ or $(\pi/2 - \varphi)$. The plot of the function $\varphi(\theta)$ for *a*=0.556 (slightly below the critical value $a_c$) is shown in fig.3a. This function is piecewise differentiable (several fractures on the plot are associated with our choice of $\varphi \in [0, \pi/2]$). One can see that the plot of $\varphi(\theta)$ approaches the axis $\varphi=0$ very closely. The minimum angle $\varphi_{\inf} = \min_{\theta \in [0,1)} \varphi(\theta)$ between the Lyapunov vectors decreases and becomes infinitely close to zero, as parameter *a* approaches the critical value $a_c$, see fig.3b. Actually it remains uncertain, whether minimum angle goes strictly to zero. However, in numerical experiment we failed to find a lower bound for the angle, which would be distinct from zero. Thus, we conjecture that the loss of smoothness of the dependences $\mathbf{k}^{1,2}(\theta)$ is associated with situations, when the "leading" and "non-leading" Lyapunov vectors $\mathbf{k}^1(\theta)$ and $\mathbf{k}^2(\theta)$ coincide at some values of the angle variable $\theta$ on torus. Note that, due to ergodicity in respect to the angle variable $\theta$, coincidence of the vectors $\mathbf{k}^1$ and $\mathbf{k}^2$ at one point of the ergodic torus implies presence of a dense set of such coincidences in images and pre-images of this point.

In order to confirm the conjecture made in the previous paragraph, let us consider the distributions of the angle $\varphi$ along typical trajectories on invariant curve for values of $a$ above the critical value $a_c$. Our interest is focused on the low bound of such distributions. Since in the numerical computations we deal with the trajectory segments of a finite length, we will observe the minimum value of the angle $\varphi$ obtained along sufficiently long segment of a typical trajectory. For the trajectory segment of $M$ iterations starting from the initial phase $\theta_0$ we define

$$\varphi_{\min}(\theta_0, M) = \min_{n=0,1,2,\ldots,M} \varphi(\theta_n).$$

Figure 4a shows a histogram of distribution of the angle $\varphi$ along segment of a typical trajectory of length $M=10^5$ on the smooth torus at $a=0.6$. The histogram shows that the probability density function is nonzero for small angles $\varphi$. In fig.4b we see an analogous histogram of angles $\varphi$ for a segment of trajectory on the SNA ($M=10^5$) at $a=0.66$. In the both cases the angle $\varphi_{\min}(\theta_0, M)$ decreases, and approaches arbitrarily close to zero as we examine longer and longer segments of the trajectory. This result does not depend upon our choice of the initial phase $\theta_0$. To show it, let us consider the maximum value of $\varphi_{\min}(\theta_0, M)$ with respect to trajectories with different initial phases $\theta_0$:

$$\Phi_M = \max_{\theta_0 \in [0,1]} \varphi_{\min}(\theta_0, M).$$

A plot of this function obtained with an ensemble of 100 trajectories on a smooth torus (at $a=0.6$) with randomly chosen initial phases $\theta_0$ is presented in the fig.5, plot 1. One can see, that for sufficiently large $M$ the function $\Phi_M$ behaves as

$$\Phi_M \sim M^{\gamma},$$

where $\gamma \simeq -1$. Thus, our conclusion concerning zero low bound for the angle $\varphi$ is valid for all or almost all trajectories on the smooth torus. Hence, we can neglect the dependence of the minimum angle $\varphi_{\min}$ upon the initial phase $\theta_0$: $\varphi_{\min} = \varphi_{\min}(M)$. Note, that the same results for the minimum angle $\varphi_{\min}$ were obtained for trajectories on SNA, as seen in the plot 2 of fig.5, at $a=0.66$.

The same properties of the distribution of the angle $\varphi$ were observed for trajectories on smooth torus and SNA at all tested values $a \in [a_c, a_f)$. In order to illustrate this statement, let us fix a length $M$ of a trajectory segment and consider dependence of the minimum angle $\varphi_{\min}$ upon the parameter $a$: $\varphi_{\min} = \varphi_{\min}(M, a)$. Figure 6 shows dependences $\varphi_{\min}(a)$ for two fixed values $M=10^4$ (a) and $M=10^5$ (b). Comparison of these plots illustrates the effect of increase of $M$. One can observe essential (in an order of magnitude) decrease of the minimum angle $\varphi_{\min}$ as $M$ grows from $10^4$ to $10^5$.

# 4. Structure of the parameter space of the quasiperiodically forced Hénon map

Let us consider now a configuration of regions of different dynamical behavior in the parameter space of the map (2). Figure 7 shows a section of the parameter space by $a - \varepsilon$ plane at fixed $b=0.5$. In order to distinguish dynamical regimes on the parameter plane, we calculate the largest non-trivial Lyapunov exponent $\sigma_1$ and the phase sensitivity exponent $\delta$ [22], which measures the sensitivity of a trajectory on attractor with respect to a variation of the angle variable $\theta$. Smooth attractors (e.g. torus T, double torus 2T, quadruplicate torus 4T, etc.) have a negative Lyapunov exponent ($\sigma_1<0$) without phase sensitivity ($\delta=0$). The symbols T, 2T, 4T, 8T below the diagram indicate intervals of the parameter $a$, in which the corresponding regimes take place at $\varepsilon=0$. The ultra-light gray tone corresponds to quasiperiodic regions characterized by smooth dependence of the Lyapunov vectors upon the angle variable on torus (the indices *1*, *3*, *6* correspond to the tori T, 2T, 4T, respectively). The light gray tone shows the regions of tori with non-smooth dependence of the Lyapunov vectors upon the angle variable (*2*, *4*, *7* correspond to T, 2T, 4T). In the regions shown in gray the Lyapunov vectors on tori are not defined (*5*, *8* correspond to 2T, 4T). The area of chaotic dynamics ($\sigma_1>0$) is shown in black. Between the regular and chaotic regimes, SNA exists in the region shown in dark-gray tone. This intermediate type of attractor is characterized by negative Lyapunov exponent ($\sigma_1<0$) with high phase sensitivity ($\delta>0$). In the white area, the system (2) has no attractors, and the trajectories escape to infinity.

Let us consider mechanisms of dynamical transitions on the parameter plane in some details. In the regions *1* and *2* the attractor of the system (2) is a smooth torus. In the region *1* this torus is characterized by a smooth dependence of Lyapunov vectors $\mathbf{k}^{1,2}(\theta)$ upon the angle $\theta$, as it shown in the figs.2b,c. In the region *2* the vector-functions $\mathbf{k}^{1,2}(\theta)$ become non-differentiable (see figs.2d,e). Transition from *1* to *2* is associated with the mechanism described in the previous section.

When crossing the line $D_1$ on the border of the regions *1* and *3*, the torus T becomes unstable and bifurcates to the double torus 2T. An example of the double torus of the map (2) at $a=0.315$, $\varepsilon=0.3$ (region *3*) is shown in fig.8a. Since the double torus 2T consists of two smooth branches:

$$2T_1 : \{(x, y, \theta) \in \mathbf{R}^2 \times \mathbf{T}^1 \mid x = x^{(1)}(\theta), y = y^{(1)}(\theta), \theta \in [0,1)\},$$
$$2T_2 : \{(x, y, \theta) \in \mathbf{R}^2 \times \mathbf{T}^1 \mid x = x^{(2)}(\theta), y = y^{(2)}(\theta), \theta \in [0,1)\},$$

we need to introduce two pairs of vector-functions $\mathbf{k}_i^{1,2}(\theta)$ ($i = 1,2$) to characterize the dependences of Lyapunov vectors upon the angle variable on the double torus. Let the pair of vector-functions $\mathbf{k}_1^{1,2}(\theta) = (k_{x1}^{1,2}(\theta), k_{y1}^{1,2}(\theta), 0)$ determine the "leading" and "non-leading" Lyapunov vectors on the branch $2T_1$, while the pair $\mathbf{k}_2^{1,2}(\theta) = (k_{x2}^{1,2}(\theta), k_{y2}^{1,2}(\theta), 0)$ be associated with the branch $2T_2$. In order to obtain numerically the value of the "leading" Lyapunov vector $\mathbf{k}_1^1(\theta_0)$ at the point $(x_0, y_0, \theta_0) \in 2T_1$, we can start iterating the map (6) from the initial angle $\theta_{-n}$ [$=\theta_0-n\omega$, mod 1], where $n$ is a sufficiently large natural number, with an arbitrarily chosen initial vector $\mathbf{k}_{-n}$. Note that the variables $x$ and $y$ in the map (6) are functions of the angle variable $\theta$, and in this case they must be defined as

$$x_j = \begin{cases} x^{(1)}(\theta_j), & j = 2m, \\ x^{(2)}(\theta_j), & j = 2m+1, \end{cases} \qquad y_j = \begin{cases} y^{(1)}(\theta_j), & j = 2m, \\ y^{(2)}(\theta_j), & j = 2m+1, \end{cases} \qquad (8)$$

where $m$ is a natural number such that $0 \le m \le n/2$. Varying $\theta_0$ within the interval $[0,1)$, we will obtain the full dependence $\mathbf{k}_1^{1}(\theta)$. On the other hand, in order to find the vector $\mathbf{k}_2^{1}(\theta_0)$ at the point $(x_0, y_0, \theta_0) \in 2T_2$, one should iterate the map (6) from the initial angle $\theta_{-n}$ with an arbitrarily chosen initial vector $\mathbf{k}_{-n}$ and with the following conditions for $x$ and $y$:

$$x_j = \begin{cases} x^{(2)}(\theta_j), & j = 2m, \\ x^{(1)}(\theta_j), & j = 2m+1, \end{cases} \qquad y_j = \begin{cases} y^{(2)}(\theta_j), & j = 2m, \\ y^{(1)}(\theta_j), & j = 2m+1. \end{cases} \qquad (9)$$

In the same way, one can determine the "non-leading" Lyapunov vector $\mathbf{k}_1^{2}(\theta_0)$ or $\mathbf{k}_2^{2}(\theta_0)$. For this purpose one should iterate the map (7) from the initial angle $\theta_n$ [$= \theta_0 + n\omega$, mod 1], where $n$ is a sufficiently large natural number, with an arbitrarily chosen initial vector $\mathbf{k}_n$, and the dependences $x=x(\theta)$ and $y=y(\theta)$ are given by the formulas (8) or (9).

The plots of the functions $k_{x1}^{1}(\theta)$ and $k_{x2}^{1}(\theta)$ at $a$=0.315, $\varepsilon$=0.3 (region **3**) are presented in fig.8b, while fig.8c shows the plots of $k_{x1}^{2}(\theta)$ and $k_{x2}^{2}(\theta)$. One can see that all these functions are smooth. Hence, the Lyapunov vectors smoothly depend upon angle $\theta$ on the double torus at the respective parameter values. On the other hand, in the region **4** the double torus is characterized by non-smooth dependence of Lyapunov vectors upon the angle variable. An example of the double torus of the map (2) at $a$=0.33, $\varepsilon$=0.3 (region **4**) is presented in fig.8d. Figs.8e,f show the plots of the non-smooth functions $k_{x1}^{1}(\theta)$ and $k_{x1}^{2}(\theta)$ at the same parameter values. In the region **5** a vicinity of double torus is of a focal type, therefore Lyapunov vectors are not defined. In this situation the attractors of the maps (6) and (7) represent a three-frequency tori.

The line $D_2$ on the border of the regions **3** and **6** (see the enlarged fragment of the parameter plane in the fig.7b) corresponds to the second doubling bifurcation, in which the double torus 2T bifurcates to the quadruplicate torus 4T. The last one is characterized by four pairs of vector-functions $\mathbf{k}_i^{1,2}(\theta)$ ($i$=1,…,4), which give the dependences of the Lyapunov vectors upon the angle variable $\theta$ on each of the four branches $4T_i$ ($i$=1,…,4) of the quadruplicate torus 4T. The regions corresponding to smoothness and non-smoothness of the vector-functions $\mathbf{k}_i^{1,2}(\theta)$ ($i$=1,…,4) are denoted as **6** and **7**, respectively. In the region **8** the Lyapunov vectors are undefined. This corresponds to the focal structure of a vicinity of the quadruplicate torus 4T.

Basing on the figure 7, one can make the following important observation: the torus-doubling bifurcations occurs when passing between the regions characterized by smooth dependence of the Lyapunov vectors on torus upon the angle variable: **1→3**, **3→6**. Indeed, the smooth dependence of Lyapunov vectors upon the angle variable on the "parent" torus is necessary for the doubling bifurcation could take place, and the "newly-born" torus is also characterized by smooth vector-functions $\mathbf{k}_i^{1,2}(\theta)$. Fig.7c shows the enlarged fragment of the parameter plane in the region where the termination of the torus-doubling line $D_1$ occurs. In order to understand the mechanism of this phenomenon, note that the line $F_1$ corresponding to the loss of smoothness of the vector-functions $\mathbf{k}^{1,2}(\theta)$ intersects with the line $D_1$ at $(a_c^{(1)}, \varepsilon_c^{(1)}) \approx (0.55478, 0.52846)$. If the parameter $\varepsilon$ is fixed at $\varepsilon < \varepsilon_c^{(1)}$, and the parameter $a$ is varied, one can observe a transition between the regions *1*→*3*. For the case $\varepsilon > \varepsilon_c^{(1)}$ such transition becomes impossible due to the loss of smoothness of the vector-function $\mathbf{k}^{1,2}(\theta)$. Note, that a vicinity of the torus-doubling terminal point contains parameter values related to the regions of different dynamical behavior: quasiperiodicity (*1*,*2*,*3*,*4*), SNA and chaos.

The line of second torus-doubling bifurcation $D_2$ terminates at $(a_c^{(2)}, \varepsilon_c^{(2)}) \approx (0.87364, 0.07471)$. Although we did not studied this phenomenon in details, we found that it is also associated with the loss of smoothness of the dependences of Lyapunov vectors upon the angle variable. For $\varepsilon < \varepsilon_c^{(2)}$ the double torus 2T may undergoes bifurcation to the quadruplicate torus 4T under variation of the control parameter $a$. For $\varepsilon > \varepsilon_c^{(2)}$ such bifurcation appears to be prevented by the loss of smoothness of the vector-functions $\mathbf{k}_i^{1,2}(\theta)$ ($i$=1,2).

On the other hand, one can see that the transitions from quasiperiodicity to SNA occur on coming out of the quasiperiodic regions characterized by non-smooth vector-functions $\mathbf{k}_i^{1,2}(\theta)$: *2*→SNA, *4*→SNA, *7*→SNA. Thus, appearance of non-smooth dependence of Lyapunov vectors upon angle variable on torus always precedes a destruction of regular quasiperiodic dynamics and onset of a strange nonchaotic attractor.

## 5. Analysis of rational approximations

Another way to explain the mechanism of termination of the torus-doubling bifurcation line is provided by the method of rational approximation, which is widely used for analysis of Hamiltonian and dissipative systems. In application to the quasiperiodically forced systems the idea of the method consists in the following (see Refs. [4-6,26,27]). The irrational parameter of frequency $\omega$ in the map (2) can be approximated by a sequence rational values $\omega_k$, such that $\omega = \lim_{k \to \infty} \omega_k$. For the case of the "golden mean" value of $\omega$ the sequence of approximants $\{\omega_k\}_{k=0,1,\ldots,\infty}$ is given by the rates of Fibonaci numbers: $\omega_k = F_{k-1}/F_k$, where $F_{k+1} = F_k + F_{k-1}$ with $F_0=0$ and $F_1=1$. For a definite level of approximation $k$, we consider an ensemble of maps

$$\begin{aligned}
x_{n+1} &= a - x_n^2 + y_n + \varepsilon \cos 2\pi \theta_n, \\
y_{n+1} &= bx_n, \\
\theta_{n+1} &= \theta_n + \omega_k, \mod 1,
\end{aligned} \qquad (10)$$

which are forced periodically with the same rational frequency $\omega_k$ and with different values of the initial angle $\theta_0$. The attractor of the map (10) depends upon the initial angle $\theta_0$. Changing $\theta_0$ continuously in the whole interval $[0, 1/F_k]$, we obtain $k$-th approximation of the attractor of the map (2) as a union of all occurring attractors of the map (10). We suppose that properties of the original system (2) can be obtained in the quasiperiodic limit at $k\to\infty$.

An approximating set of order $k$ for attracting torus represents a smooth set of stable periodic orbits of period $F_k$. Note, that the approximating orbits may be of two types: node and focus. Let us consider a periodic orbit of the map (10) that starts from the initial angle $\theta_0$: $(x_0, y_0, \theta_0)$, $(x_1, y_1, \theta_1)$, …, $(x_{F_k-1}, y_{F_k-1}, \theta_{F_k-1})$. The monodromy matrix of the periodic orbit is:

$$\hat{J}^{(F_k)}(x_0, y_0, \theta_0) = \hat{J}(x_{F_k-1}, y_{F_k-1}, \theta_{F_k-1})\hat{J}(x_{F_k-2}, y_{F_k-2}, \theta_{F_k-2})\ldots\hat{J}(x_0, y_0, \theta_0). \quad (11)$$

Since the given orbit belongs to a smooth approximating set, the variables $x$ and $y$ in (11) are functions of the angle variable $\theta$ ($x_0 = x(\theta_0)$, $y_0 = y(\theta_0)$, etc.), and we can write simply $\hat{J}^{(F_k)}(\theta_0)$. The type of the periodic orbit is determined by the values of the multipliers $\mu_{1,2}$, which represent the nontrivial eigenvalues of the matrix $\hat{J}^{(F_k)}(\theta_0)$. The multipliers depend upon the initial angle of the orbit $\theta_0$: $\mu_{1,2} = \mu_{1,2}(\theta_0)$. If the multipliers are real, the orbit is of a nodal type, otherwise it is a focus. The Lyapunov exponents of the orbit are defined as

$$\sigma_{1,2}(\theta_0) = (1/F_k)\ln|\mu^{1,2}(\theta_0)|.$$

In the same way, the "leading" and "non-leading" Lyapunov vectors $\mathbf{k}^{1,2}(\theta_0)$ can be defined as eigenvectors of the matrix $\hat{J}^{(F_k)}(\theta_0)$:

$$\hat{J}^{(F_k)}(\theta_0)\,\mathbf{k}^{1,2}(\theta_0) = \mu_{1,2}(\theta_0)\,\mathbf{k}^{1,2}(\theta_0),$$

and they depend upon the initial angle $\theta_0$. Let us analyze structure of the approximating set of periodic orbits, which correspond to different values of the initial angle $\theta_0$. In the quasiperiodic limit ($k\to\infty$) we find the following 3 cases to be possible:

(C1) As the value of $\theta_0$ is varied, one can observe transition of the multipliers $\mu_{1,2}(\theta_0)$ from real to complex-conjugate values. Thus, the approximating set includes periodic orbits of two types: nodal and focal. Such "mixed" structure of the approximating set of orbits persists as the order of approximation $k$ is increased.

(C2) The multipliers $\mu_{1,2}(\theta_0)$ are real for all $\theta_0$ from the interval $[0, 1/F_k)$. Thus, the approximating set consists of periodic orbits of nodal type. However, for some values of $\theta_0$ the condition holds $|\mu_1| \geq |\mu_2|$, while for other values of $\theta_0$ the opposite is valid, $|\mu_2| > |\mu_1|$. In other words, this set has non-homogeneous structure in the sense that "leading" Lyapunov vector may transform into "non-leading" one and back, as the value of $\theta_0$ is varied. Such "non-uniform" structure of the approximating set of orbits persists as the order $k$ of the approximation is increased.

(C3) The approximating set for smooth torus consists of periodic orbits of the same type (nodal). The set has uniform structure in the sense of absence of the exchange of the "leading" and "non-leading" Lyapunov vectors.

In the cases (C1) and (C2) the structure of approximating set can be referred to as "phase-dependent". In the quasiperiodic limit, the phase-dependent approximating set forms a torus which is characterized by a non-smooth dependence of the Lyapunov vectors $\mathbf{k}^{1,2}(\theta)$ upon the angle variable $\theta$. On the other hand, the case (C3) corresponds to a situation when the dependences $\mathbf{k}^{1,2}(\theta)$ are smooth.

Before a study of the structure of approximating set, note, that the one time iterated map (10) has negative Jacobi determinant ($J = -b$). The superposition of $F_k$ maps (10) will possess the Jacobian $J=(-b)^{F_k}$. Hence, we need to consider separately rational approximants $\omega_k=F_{k-1}/F_k$ with odd and even periods $F_k$. Indeed, the matrix $\hat{J}^{(F_k)}(\theta_0)$ has the form

$$\hat{J}^{(F_k)}(\theta_0) = \begin{bmatrix} J_{11}(\theta_0) & J_{12}(\theta_0) & J_{12}(\theta_0) \\ J_{21}(\theta_0) & J_{22}(\theta_0) & J_{23}(\theta_0) \\ 0 & 0 & 1 \end{bmatrix}.$$

The nontrivial multipliers of the periodic orbit $\mu_{1,2}(\theta_0)$ are defined as

$$\mu_{1,2}(\theta_0) = S(\theta_0)/2 \pm \sqrt{(S(\theta_0)/2)^2 - (-b)^{F_k}},$$

where $S(\theta_0) = J_{11}(\theta_0) + J_{22}(\theta_0)$. Since we have originally chosen $b>0$, the multipliers are always real for the case of odd values of $F_k$. Hence, all the approximating orbits of odd period are nodal. On the other hand, for even $F_k$, such values of the angle $\theta_0$ can exist, that a condition holds

$$(S(\theta_0)/2)^2 < b^{F_k}. \tag{12}$$

For this case, the approximating set can possess orbits of both nodal and focal type.

As an example, let us consider the system (2) at the parameter values $a=0.34$, $\varepsilon=0.6$, $b=0.5$, which correspond to the case of existence of the non-smooth dependences $\mathbf{k}^{1,2}(\theta)$. In order to illustrate existence of phase-dependent structure of the approximating set of periodic orbits, we have computed the nontrivial Lyapunov exponents $\sigma_{1,2}$ as functions of the initial angle variable $\theta_0$ within the interval $[0,1/F_k)$. Figure 9a shows plots of the functions $\sigma_{1,2}(\theta_0)$ for the odd period of approximation: $F_k=55$. The exponent $\sigma_1$ corresponds to the negative multiplier ($\mu_1<0$), while the exponent $\sigma_2$ [$=\ln|b| - \sigma_1$] corresponds to the positive multiplier ($\mu_2>0$). One can see that the interval $[0,1/F_k)$ turns out to be subdivided into 3 segments: *A*, *B* and *C*. Within the subintervals *A* and *C* the condition holds $-\infty < \sigma_2 < (1/2)\ln|b| < \sigma_1 < 0$ ($0 < \mu_2 < b^{F_k/2} < -\mu_1 < 1$, respectively). Thus, the Lyapunov vector $\mathbf{k}^1$ is "leading" within these subintervals. In the subinterval *B* the backward condition holds $-\infty < \sigma_1 < (1/2)\ln|b| < \sigma_2 < 0$ ($0 < \mu_1 < b^{F_k/2} < -\mu_2 < 1$). Hence, the Lyapunov vector $\mathbf{k}^2$ appears to be "leading" in the subinterval *B*. Note, that on the border of the intervals there are two of such points $\theta^*_{1,2}$ that $\mu_2(\theta^*_{1,2}) = -\mu_1(\theta^*_{1,2}) = b^{F_k/2}$. We have tested rational approximants with large odd periods $F_k$ up to $F_k$= 4181 and found that the structure of the interval $[0, 1/F_k)$ remains qualitatively the same as the level $k$ of rational approximant increases. However, the quantitative features of the interval structure may change with $k$. Let us denote the relative lengths of the subintervals *A*, *B* and *C* as $p_A$, $p_B$, and $p_C$, respectively (note, that the sum length of all subintervals is normalized to unity: $p_A + p_B + p_C = 1$). Figure 9b shows the dependence of the relative length $p_{A+C}$ [$=p_A+p_C$] upon $k$. One can see that the dependence has irregular character. Note, that none of the two components ($p_{A+C}$ and $p_B$) decays to zero as the level $k$ increases.

Figure 9c shows the dependences of the Lyapunov exponents $\sigma_{1,2}$ upon the initial angle $\theta_0$ for the approximating set of periodic orbits in the case of even period $F_k$ = 34. In this figure the interval $[0,1/F_k)$ is divided into 5 subintervals. Within the subintervals *A*, *C* and *E* the values of multipliers $\mu_{1,2}$ are real. Hence, the corresponding stable periodic orbits are of nodal type. On the other hand, within subintervals *B* and *D* the condition (12) holds. The periodic orbits within subintervals *B* and *D* are characterized by complex-conjugate multipliers ($\mu_1 = \mu^*_2$). Transition from the region *A* to the region *B* implies a change of the nodal type of orbit to the focal type. Analyzing the structure of the interval $[0,1/F_k)$ under increase of $k$, we found that it remains qualitatively the same for large even $F_k$. However, the quantitative features change with $k$. The sum length of the intervals of nodal orbits $p_{A+C+E}$ [$= p_A + p_C + p_E$] dominates over the sum length of the intervals of focal orbits $p_{B+D}$ [$=p_B + p_D$]. In fig.9d we have plotted the sum length $p_{B+D}$ (double logarithmic scale) vs. the period $F_k$ (logarithmic scale). Since the condition (12) is applicable to both even and odd approximations, we consider both even and odd periods $F_k$ in order to obtain a representative plot. One can see that the points on the plot can be fitted by a straight line for sufficiently large $F_k$. Hence, the relative length $p_{B+D}$ decays as exponent of the period $F_k$ of approximation.

If approximating sets of some torus possess phase-dependent non-uniformity of the described type, and this non-uniformity persists as the order of approximation $k$ tends to infinity (quasiperiodic limit), the corresponding torus can not undergo a doubling bifurcation. Indeed, let us consider the mechanism of doubling bifurcation of torus from the viewpoint of bifurcations of the approximating periodic orbits. For the case of approximation with odd period $F_k$, each periodic orbit of the set undergoes doubling bifurcation along the "leading" Lyapunov direction, which is associated with the negative multiplier $\mu_2$. However, the Lyapunov vector $\mathbf{k}^1$ appears to be "leading" only within segments *A* and *C* of the interval $[0,1/F_k)$, as we have shown above. Within the segment *B* the condition holds $-b^{F_k/2} < \mu_2 < 0$. Hence, the periodic orbits within the segment *B* can not undergo doubling bifurcation. For the case of approximation with even period $F_k$, there are two subintervals (*B* and *D*), in which the periodic orbits are characterized by complex-conjugate multipliers. Obviously, the respective periodic orbits can not undergo doubling bifurcation.

Note, that arising of the phase-dependent non-uniformity in the structure of the approximating set makes impossible other regular torus bifurcations (symmetry breaking, saddle-node) besides the torus-doubling bifurcation. Indeed, the regular torus bifurcation occurs when all orbits corresponding to different initial angles $\theta_0$ on torus bifurcate in a similar way. The last requirement is obviously impossible for the case of phase-dependent non-uniform structure of the approximating set described above. Hence, the given torus can undergo evolution and destruction according to phase-dependent mechanisms only. In other words, under variation of the parameters of the map (2) such torus disappears with arising of a strange non-chaotic attractor or divergence of trajectories. The last conclusion is conformed with the numerical observations made in *Section 2* on the structure of the parameter space of the quasiperiodically forced Hénon map.

## 5. Conclusion

In the present paper we have observed a new transition which consists in appearance of non-smooth dependences of the Lyapunov vectors upon the angle variable on torus in the quasiperiodically forced Hénon map. We have argued that such transition terminates the line of torus-doubling bifurcation on the parameter plane of the model map and restricts the number of torus-doubling bifurcations on the route to chaos. The new transition is shown to precede the destruction of a regular quasiperiodic motion and the birth of a strange nonchaotic attractor in the model map.

We suppose that arguments of this paper concerning with the mechanism of torus-doubling bifurcation can be applied as well to other regular torus bifurcations, namely symmetry breaking, transcritical and saddle-node bifurcations. One can show that existence of smooth dependence of the Lyapunov vectors upon the angle variable on torus is a necessary condition for possibility of these regular torus bifurcations. Therefore, we suppose that the new phenomenon which consists in appearance of a non-smooth dependence of Lyapunov vectors upon the angle variable on torus plays a key role in different scenarios of transition from regular motion to chaos in the quasiperiodically forced systems.

## Acknowledgements

The work was supported by the Russian Foundation of Basic Research (grant No. 03-02-16074) and by the grant of the UK Royal Society. The authors thank S.P. Kuznetsov for useful discussions.

Figures:

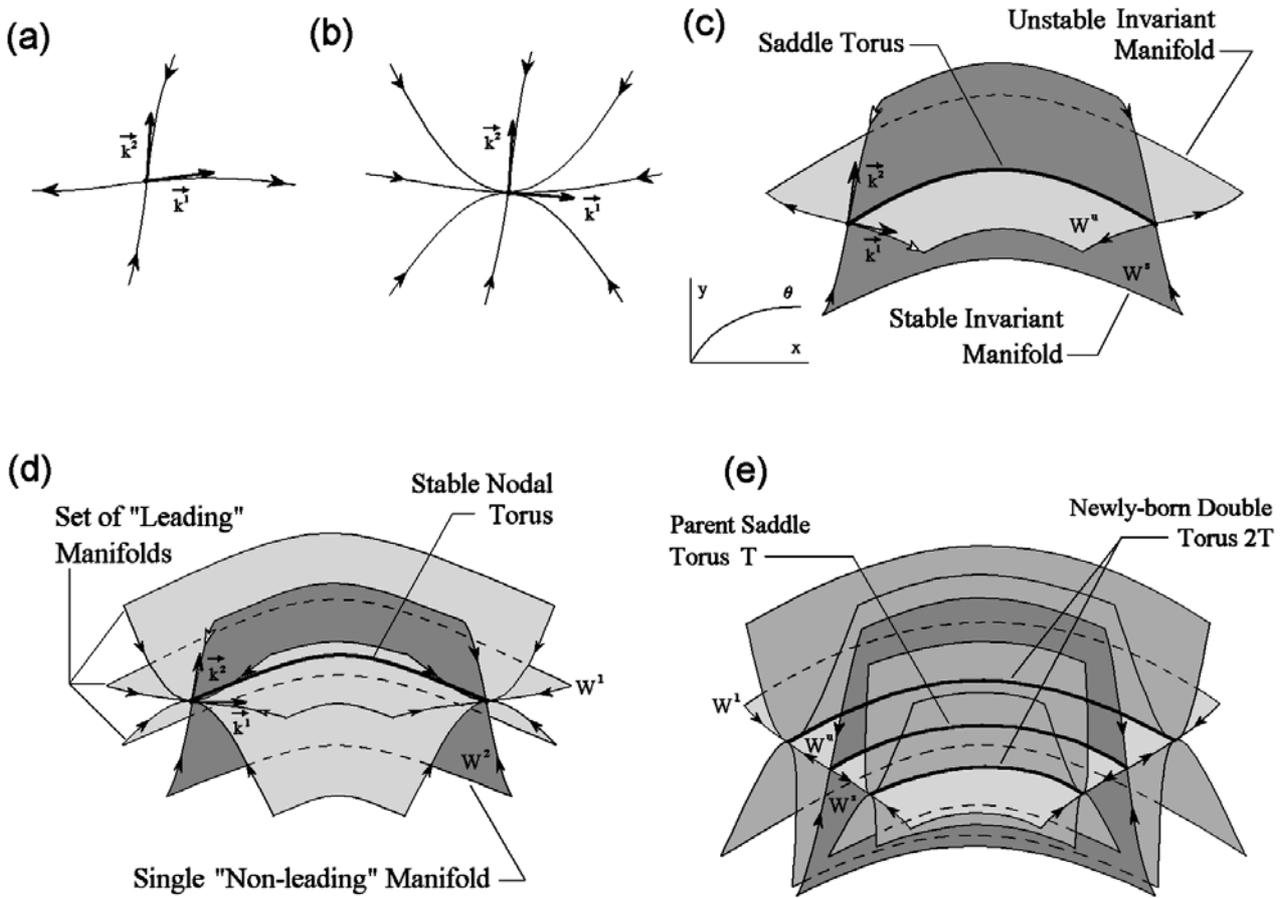

**Fig.1.** Schematic drawings of the fixed points, tori and associated invariant manifolds. (a) Saddle fixed point of the map (1). (b) Nodal fixed point of the map (1). (c) Saddle torus of the map (2). (d) Nodal torus of the map (2). (e) Parent saddle torus T and the newly-born double torus 2T. The detailed explanations are provided in *Section 2* of the paper.

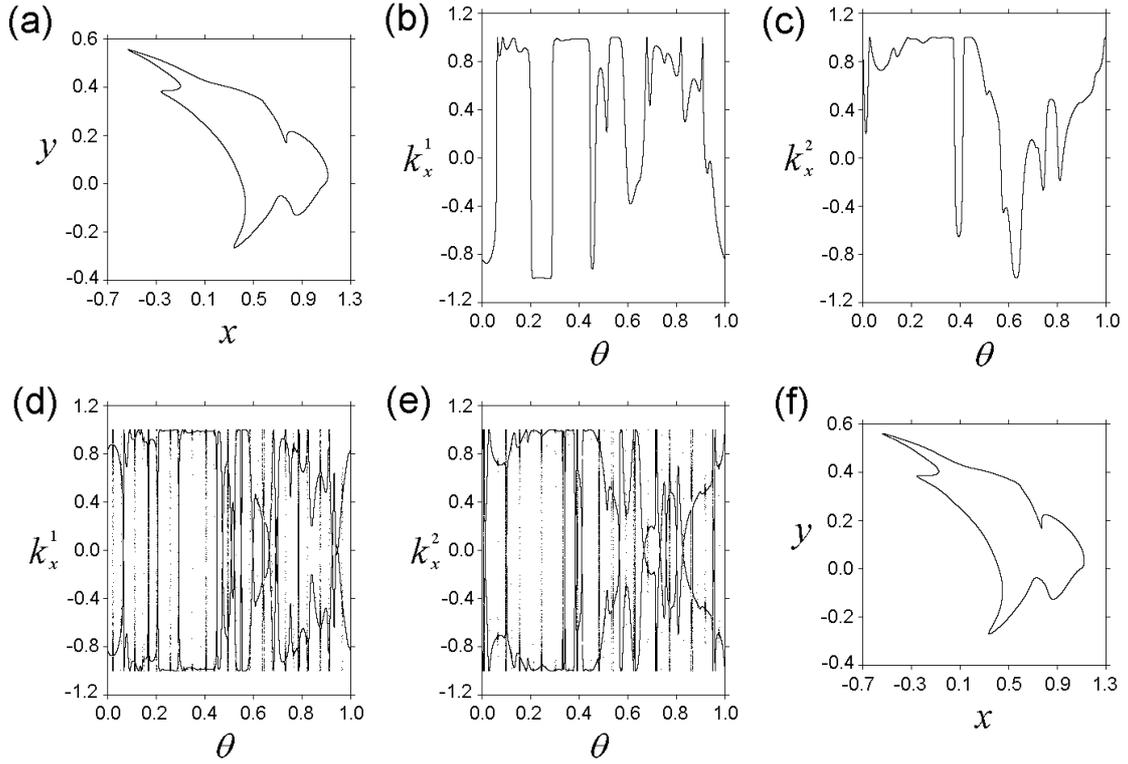

**Fig.2.** (a) Attracting torus of the map (2) at $a=0.55$, $\varepsilon=0.6$. (b) The plot of the function $k_x^1(\theta)$ at $a=0.55$, $\varepsilon=0.6$ (only odd iterations of the map (6) are plotted). (c) The plot of the function $k_x^2(\theta)$ at $a=0.55$, $\varepsilon=0.6$ (image of the torus of the map (7)). (d) SNA of the map (6) at $a=0.559$, $\varepsilon=0.6$. (e) SNA of the map (7) at $a=0.559$, $\varepsilon=0.6$. (f) Attracting torus of the map (2) at $a=0.559$. We have chosen $b=0.5$ for these and all the following figures.

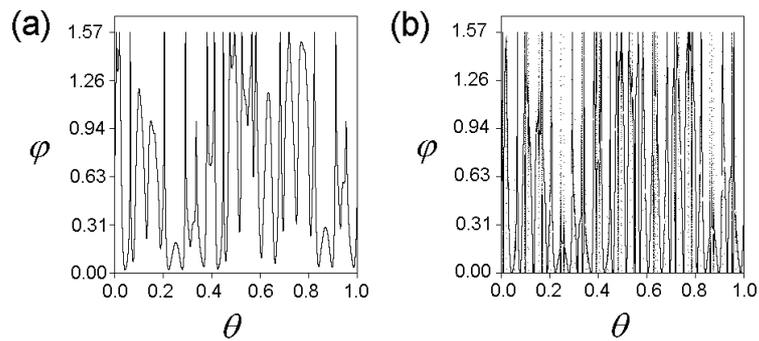

**Fig.3.** The dependences of the angle $\varphi$ between Lyapunov vectors upon the angle coordinate $\theta$: (a) at $a=0.556$ (slightly below the critical value $a_c$), $\varepsilon=0.6$, (b) at $a=0.559$ (slightly above the critical value $a_c$), $\varepsilon=0.6$.

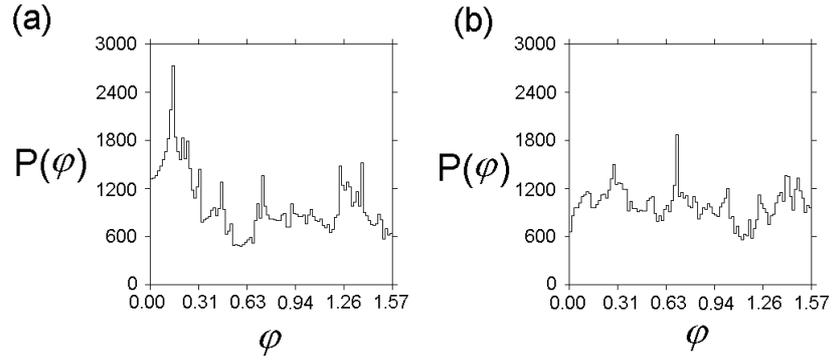

**Fig.4:** Histograms of the angle $\varphi$ between Lyapunov vectors: (a) for trajectory on the torus at $a=0.6$, $\varepsilon=0.6$, (b) for trajectory on SNA at $a=0.66$, $\varepsilon=0.6$. The length of trajectory segment is $10^5$ iterations.

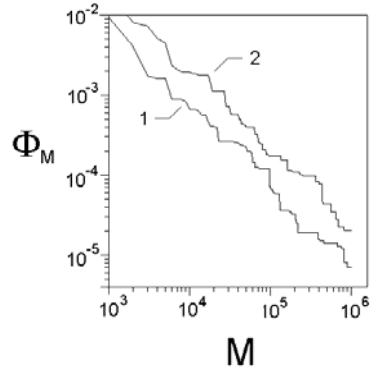

**Fig.5:** The plot of the function $\Phi_M$ for 100 trajectories with randomly chosen initial angle variable on torus (plot 1, $a=0.6$, $\varepsilon=0.6$) and on the SNA (plot 2, $a=0.66$, $\varepsilon=0.6$).

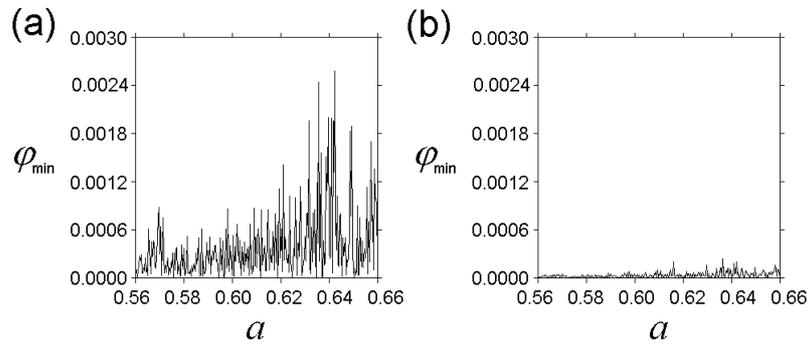

**Fig.6:** The effect of increase of the trajectory segment length $M$ upon the minimum angle between Lyapunov vectors: dependence $\varphi_{min}(M, a)$ versus $a$ for (a) $M=10^4$ and (b) $M=10^5$.

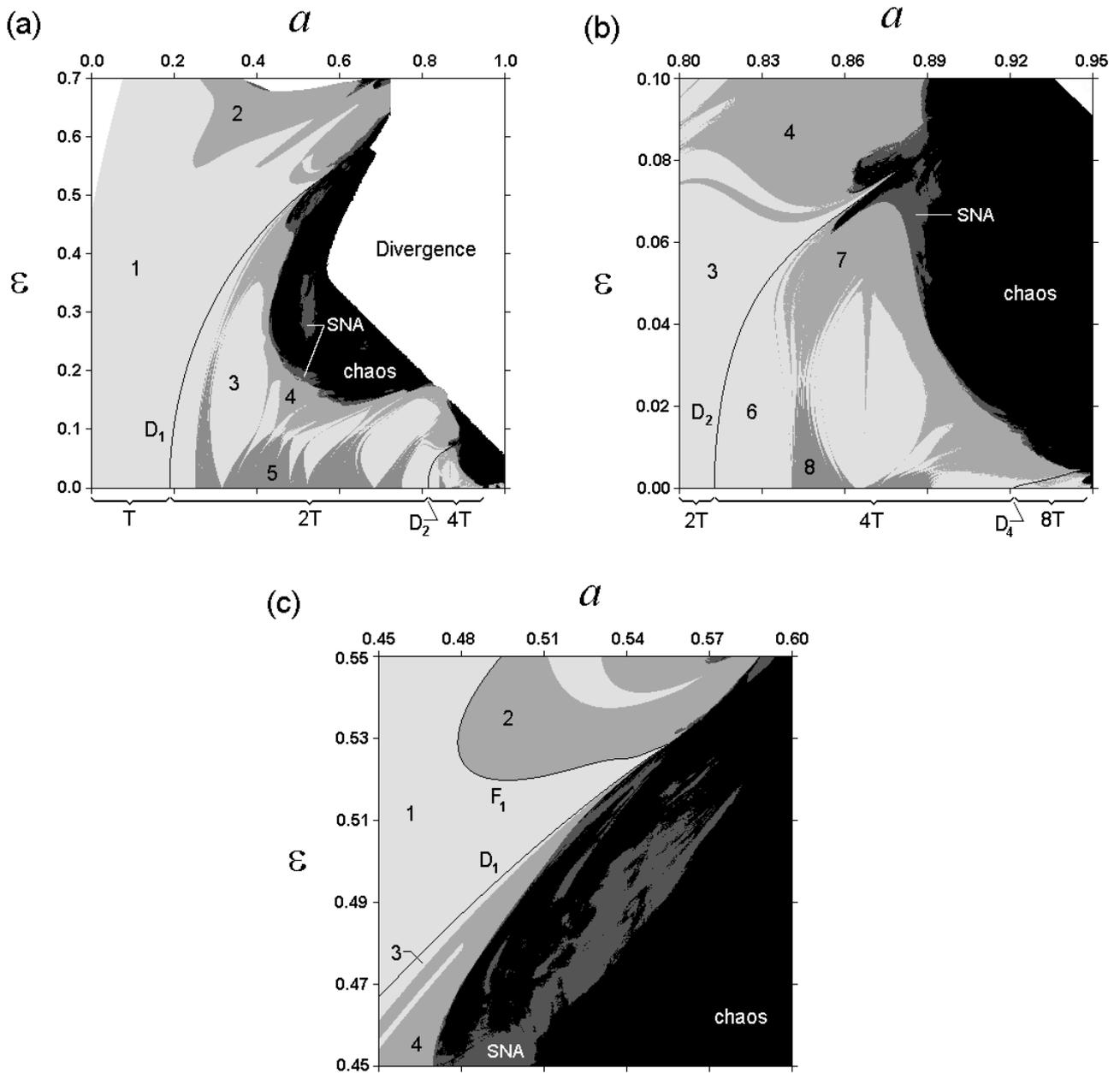

**Fig.7:** (a) Parameter plane of the map (2) the at fixed value $b$=0.5. The regions of existence of torus T (*1,2*), double torus 2T (*3,4,5*) and quadruplicate torus 4T (*6,7,8*) are divided into sub-regions in accordance with smooth, non-smooth or undefined dependences of the Lyapunov vectors upon the angle variable. The ultra-light gray (*1,3,6*), light gray (*2,4,7*) and gray (*5,8*) tones denote the regions of smooth, non-smooth or undefined dependences, respectively. The regions of SNA, chaos and divergence are shown in dark gray, black and white, respectively. The lines $D_1$ and $D_2$ represent the first and the second torus-doubling bifurcations curves. The line $F_1$ on the border of the regions *1* and *2* denotes the curve of the loss of smoothness of the vector-function $\mathbf{k}^{1,2}(\theta)$. (b), (c) The enlarged fragments of the parameter plane.

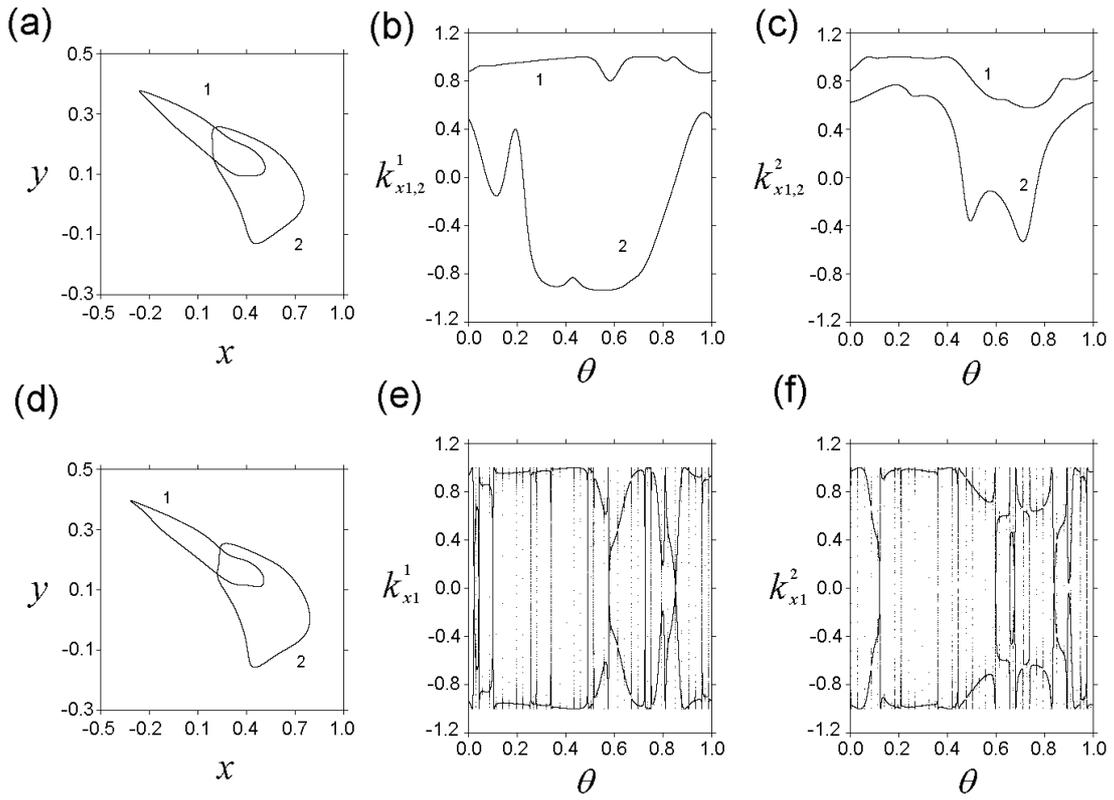

**Fig.8.** (a) Attracting double torus of the map (2) at $a$=0.315, $\varepsilon$=0.3 (digits *1* and *2* are related to the branches $2T_1$ and $2T_2$ ). (b) The plots of the functions $k_{x1}^1(\theta)$ and $k_{x2}^1(\theta)$ at $a$=0.315, $\varepsilon$=0.3. (c) The plots of the functions $k_{x1}^2(\theta)$ and $k_{x2}^2(\theta)$ at $a$=0.315, $\varepsilon$=0.3. (d) Attracting double torus of the map (2) at $a$=0.33, $\varepsilon$=0.3. (e) The plot of the function $k_{x1}^1(\theta)$ at $a$=0.33, $\varepsilon$=0.3. (c) The plot of the function $k_{x1}^2(\theta)$ at $a$=0.33, $\varepsilon$=0.3.

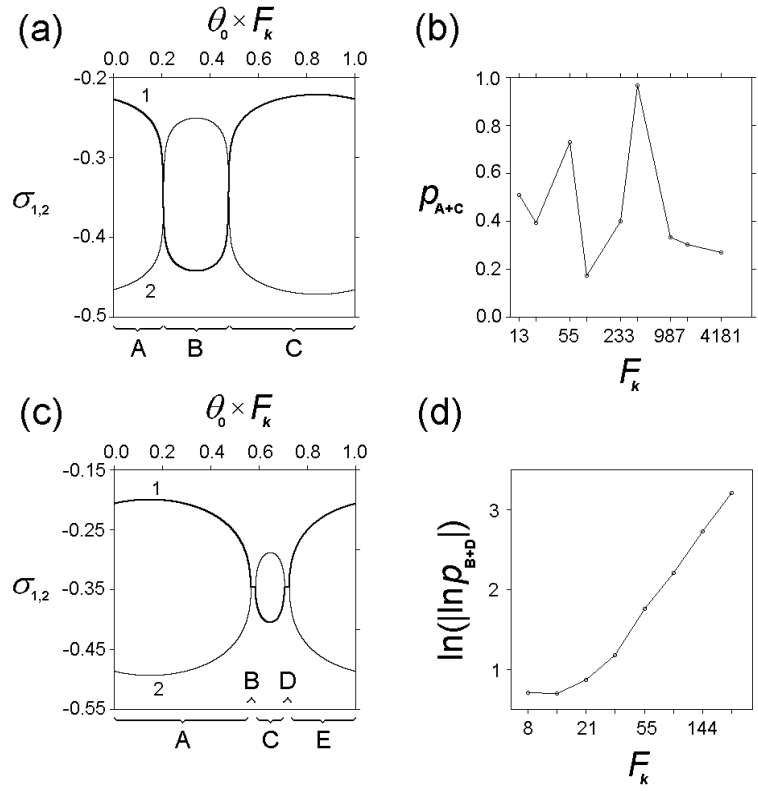

**Fig.9:** (a) The plots of the functions $\sigma_1(\theta_0)$ (thick curve) and $\sigma_2(\theta_0)$ (thin curve) for $F_k=55$ at $a=0.34$, $\varepsilon=0.6$. (b) The dependence of the sum length $p_{A+C}$ upon the period $F_k$ (logarithmic scale) of approximation. (c) The plots of the functions $\sigma_1(\theta_0)$ (thick curve) and $\sigma_2(\theta_0)$ (thin curve) for $F_k=34$ at $a=0.34$, $\varepsilon=0.6$. (d) The dependence of the sum length $p_{B+D}$ (double logarithmic scale) upon the period $F_k$ (logarithmic scale) of approximation.